\documentclass[preprint]{elsarticle}
\usepackage{bm}
\usepackage{amsmath}
\usepackage{epsfig}
\usepackage{rotating}
\biboptions{sort&compress}
\usepackage{graphicx}        
\usepackage{lipsum}

\def\pr{\prime}
\def\be{\begin{equation}}
\def\lan{\left\langle}
\def\ran{\right\rangle}
\def\ee{\end{equation}}
\def\barr{\begin{array}}
\def\earr{\end{array}}

\def\l{\left}
\def\r{\right}
\def\dis{\displaystyle}
\def\ed{\end{document}}
\def\f{\frac}

\def\co{{\cal O}}

\def\ce{{\cal E}}
\def\cam{{\cal M}}

\def\cs{{\bf s}}

\def\eh{{\hat{E}}}
\def\ehh{{\hat{H}}}

\def\wm{{\widetilde {m}}}

\def\tmp{\widetilde{m_p}}
\def\tmn{\widetilde{m_n}}
\def\wm{{\widetilde {m}}}
\def\ed{\end{document}}

\begin{document}

\begin{frontmatter}

\title{Statistical Nuclear Spectroscopy with $q$-normal and bivariate $q$-normal distributions and $q$-Hermite polynomials}

\author[add1]{V. K. B. Kota}
\ead{vkbkota@prl.res.in}
\author[add2]{Manan Vyas \corref{cor1}}
\ead{manan@icf.unam.mx}

\cortext[cor1]{Corresponding author}
\address[add1]{Physical Research Laboratory, Ahmedabad 380 009, India}
\address[add2]{Instituto de Ciencias F{\'i}sicas, Universidad Nacional Aut{\'o}noma de M\'{e}xico, 62210 Cuernavaca, 
M\'{e}xico}

\begin{abstract}

Statistical nuclear spectroscopy (also called spectral distribution method), introduced by J.B. French in late 60's and developed in detail in the later years by his group and many other groups, is based on the Gaussian forms for the state (eigenvalue) and transition strength densities in shell model spaces with their extension to partial densities defined over shell model subspaces. The Gaussian forms have their basis in embedded random matrix ensembles with nuclear Hamiltonians consisting of a mean-field one-body part and a residual two-body part. However, following the recent random matrix results for the so called Sachdev-Ye-Kitaev model due to Verbaaarschot et al, embedded random matrix ensembles with $k$-body interactions are re-examined and it is shown that the density of states, transition strength densities and strength functions (partial densities) in fact follow more closely the $q$-normal distribution (the parameter $q$ is related to the fourth moment of these distributions with $q=1$ giving Gaussian and $q=0$ giving semi-circle form). The $q$-normal has the important property that it is bounded for $0 \le q < 1$. The $q$-normal (also its bivariate and general multi-variate extensions) and the associated $q$-Hermite polynomials are studied for their properties by Bryc, Szabowski and others [P.J. Szabowski, Electronic Journal of Probability {\bf 15}, 1296 (2010)]. Following these, in the present article, developed is statistical nuclear spectroscopy based on $q$-normal (univariate and bivariate) distributions and the associated $q$-Hermite polynomials. In particular, formulation is presented for nuclear level densities, shell model orbit occupancies, transition strengths (for electromagnetic and $\beta$ and double $\beta$-decay type operators) and strength sums. 

\end{abstract}


\end{frontmatter}

\section{Introduction}

Statistical nuclear spectroscopy or spectral distribution method (SDM) is based on the Gaussian form of the state (eigenvalue) densities and transition strength densities in shell model spaces. Construction of these distributions, without recourse to shell model matrix diagonalizations, allow one to calculate observables such as level densities, orbit occupancies, $\beta$-decay rates for astrophysical applications and so on. This subject was originally introduced by J.B. French \cite{jbf1,jbf2, jbf3,jbf4,DFW,DFW2} and developed much further in the last 50 years by the Rochester group and many other groups. Some early reviews are in \cite{MM,FK,KK} and an early book is due to S.S.M. Wong \cite{Wong}. Many of the important papers on the subject till 2006 are reprinted in a book by Kota and Haq along with a long introduction to the subject of spectral distributions in nuclei \cite{KH}. Significant developments in this subject in the last 10 years are due to Zelevinsky, Horoi and Se'nkov\cite{Zel1,Zel2,Zel3,Zel4,Zel5}. There are now significant applications of SDM to nuclear level densities, binding energies of neutron rich $sd$-shell nuclei, shell model orbit occupancies for nuclei of interest in double $\beta$-decay, $\beta$-decay rates for presupernovae evolution, double $\beta$-decay transition matrix elements, goodness of group symmetries, analysis of operators and so on; see\cite{KH,Go-12,KoChv,Zel5,Jcpx1,Jcpx2,Quen1,SDV} and references therein. In addition, there are also applications in atomic and molecular physics \cite{app1,app2,app3,app4,app5,app6}.

Many-body quantum chaos and random matrix theory provide the basis for statistical nuclear spectroscopy. Classical random matrix ensembles, i.e. the Gaussian orthogonal, unitary and symplectic ensembles (GOE, GUE and GSE) are well known now in physics and need no introduction \cite{Porter,Mehta,RMT-book}. Nuclear shell model Hamiltonians ($H$) consist of a mean-field one-body part and a residual two-body interaction. With the two-body part sufficiently strong, nuclear levels in general exhibit quantum chaos and the appropriate random matrix ensembles for describing this in nuclear structure are the so-called embedded ensembles generated by $k$-body interactions [EE($k$)] in many-particle ($m$-particle with $m >k$) spaces. In particular, the embedded Gaussian orthogonal ensemble of 2-body or $(1+2)$-body interactions [EGOE(2) or EGOE(1+2)] with or without various other quantum numbers are appropriate. These ensembles are first recognized by French, Bohigas, Wong and Flores \cite{PLB1,PLB2} and that they generate Gaussian eigenvalue densities was shown analytically, using the so called 'binary correlation approximation', by Mon and French \cite{MF}. For the definition of these ensembles with general $k$-body interactions (note that for $m=k$, EE will reduce to the classical ensembles giving the well known Wigner semi-circle form for the eigenvalue density), their various Lie-algebraic and other extensions and the results for the statistical properties generated by them see \cite{MF,Br-81,Fl-Iz,Ko-01,BRW,EE-su4,Ko-book,SM,BISZ,CK} and references therein. For the EE basis for SDM see also \cite{FKPT,Ko-03,KoChv}.

Recently Verbaarschot and collaborators started analyzing quantum chaos in the Sachdev-Ye-Kitaev (SYK) model using random matrix theory \cite{Verb1,Verb2,Verb3,Verb4,Verb5}. Most significant result in these papers, that is relevant for SDM, is the recognition that the $q$-normal distribution, mentioned first by Bozejko et al \cite{Boze} in the context of some non-commutative models, indeed gives the eigenvalue density in the SYK model. Bryc, Szablowski, Ismail and others later clearly showed that this $q$-normal distribution (see Section \ref{sec2} for definition and other mathematical details) has a purely commutative and classical probabilistic meaning. Similar is the situation with the associated $q$-Hermite and $q$-Al-Salam-Chihara polynomials
\cite{Bryc1,Bryc2,Sza-1,Sza-2,Ismail}. All the $q$ distributions and polynomials used in this paper are defined in Section \ref{sec2} ahead. With the $q$-normal reducing to Gaussian form for $q=1$ and semi-circle form for $q=0$, immediately shows that EE($k$) perhaps will generate $q$-normal form for the eigenvalue densities. This possibility was investigated recently by using the formulas for the lower order moments generated by EGOE($k$) and EGUE($k)$. Remarkably, it is seen that the lower order moments (up to 8th order) of the eigenvalue density generated by EE($k$) are essentially identical to the lower order moments given by $q$-normal distribution \cite{qMK-1} with the fourth moment determining the value of the $q$ parameter. Similarly, it is shown that the lower order bivariate moments (verified up to order 6) of the transition strength densities (see Section \ref{sec5} for definition) generated by EE are indeed essentially same as those of the bivariate $q$-normal distribution \cite{qMK-2}. Going further, it is also seen that the lower order moments of strength functions or local density of states (also called partial densities) generated by EE are close to those from the conditional $q$-normal distribution \cite{qMK-3}. All these results are also supported by several numerical calculations using EE for both fermion and boson systems; see \cite{qMK-1,qMK-2,qMK-3,chv1}. An important property of the $q$ distributions is that they are bounded unlike a Gaussian. The necessity to introduce a cut-off in the tails of the Gaussians is seen in several studies in the past \cite{Zel3,Zel4,KH,Lea}. Following all these, it is clear that it is important to develop SDM using $q$-normal  distributions and its relatives so that the information about the fourth moment of the density of eigenvalues and the boundedness of the eigenvalue density and other distributions are in a natural way incorporated in SDM. The purpose of the present paper is to address this by giving the basic approaches one may adopt using $q$-normal distributions and the associated $q$-Hermite polynomials in SDM. Now we will give a preview.  

In Section \ref{sec2}, the functions - $q$-normal $f_{qN}$, bivariate $q$-normal $f_{biv-qN}$ and the conditional $q$-normal $f_{CqN}$ are given along with $q$-Hermite polynomials. Collected also are some of their properties that are used in the later Sections. In Section \ref{sec3}, formulation for calculating nuclear state densities using $q$-normal distribution is described. Discussed briefly also are (i) determination of the ground state energy and (ii) angular momentum $J$ projection from state densities giving level densities. Section \ref{sec4} gives methods for obtaining shell model orbit occupancies using $q$-normal form and expansions using $q$-Hermite polynomials. In Sections \ref{sec3} and \ref{sec4}, conditional $q$-normal distribution also plays a role. Section \ref{sec5} gives the formulation for transition strengths and the associated sum rule quantities using bivariate $q$-normal distribution. Finally, Section \ref{sec6} gives conclusions and future outlook.  

\section{$q$-normal distribution, lower order moments and $q$-Hermite polynomials}
\label{sec2}

Given a normalized probability distribution $\rho(z)$, its moments are $M_p$ defined by
\be
M_P = \dis\int z^p \,\rho(z)\, dz
\label{eq.sdm1}
\ee
with the centroid $\epsilon=M_1$ and the variance $\sigma^2=M_2-(M_1)^2$ (note that $\sigma$ is called width). The first moment or the centroid $\epsilon$ gives the location and the second moment or width gives the scale for the distribution $\rho$. A standardized variable $x=(z-\epsilon)/\sigma$ is zero centered with variance unity. Then, $\rho(z)dz = \eta(x)dx$ with $\epsilon=0$ and $\sigma=1$ for $\eta(x)$. Now, the moments 
\be
\mu_p = \dis\int x^p\,\eta(x)\,dx
\label{eq.sdm2}
\ee
of $\eta(x)$ are reduced central moments (note that $\mu_1=0$ and $\mu_2=1$). The higher reduced central moments $\mu_p$ with $p \ge 3$ define the shape of $\rho$. The third moment $\mu_3=\gamma_1$ is the skewness and the fourth moment $\mu_4$ defines excess or kurtosis $\gamma_2=\mu_4-3$. These are the most important shape parameters. Just as above, for bivariate distributions we can define bivariate moments and bivariate reduced central moments \cite{Kendall,KH}. For symmetrical distributions, $\mu_p=0$ for $p$ odd. 

In addition to the moments, let us also introduce $q$ numbers $[n]_q$ defined by
\be
\l[n\r]_q = \dis\frac{1-q^n}{1-q} = 1+q + q^2 + \ldots+q^{n-1}\;.
\label{eq.sdm3}
\ee
Note that $[n]_{q \rightarrow1}=n$. Similarly $[n]_q! = \dis\Pi^{n}_{j=1} \,[j]_q$ with $[0]_q!=1$.

\subsection{$q$-normal distribution and $q$-Hermite polynomials}
\label{q-nrm}

Let us begin with the $q$-normal distribution $f_{qN}(x|q)$ \cite{Ismail,Sza-1}, with $x$ being a standardized variable (then $x$ is zero centered with variance unity),
\be
f_{qN}(x|q) = \dis\frac{\dis\sqrt{1-q} \dis\prod_{k^\pr=0}^{\infty} \l(1-
q^{k^\pr +1}\r)}{2\pi\,\dis\sqrt{4-(1-q)x^2}}\; \dis\prod_{k^\pr=0}^{\infty}
\l[(1+q^{k^\pr})^2 - (1-q) q^{k^\pr} x^2\r]\;.
\label{eq.sdm4}
\ee
The $f_{qN}(x|q)$ is defined over $S(q)$ with
\be
S(q) = \l(-\dis\frac{2}{\dis\sqrt{1-q}}\;,\;+\dis\frac{2}{\dis\sqrt{1-q}}\r)
\label{eq.sdm5}
\ee
and $q$ takes values $0$ to $1$ (in this paper). Note that $f_{qN}(x|q) = 0$ outside $S(q)$ and the integral of $f_{qN}(x|q)$ over $S(q)$ is unity,
$$
\dis\int_{S(q)} f_{qN}(x|q)\,dx =1\;.
$$
For $q=1$ taking the limit properly
will give $f_{qN}(x|1)= (1/\sqrt{2\pi})\,\exp-x^2/2$, the Gaussian with $S(q=1)=(-\infty , \infty)$. Also, 
$f_{qN}(x|0)=(1/2\pi) \sqrt{4-x^2}$, the semi-circle with $S(q)=(-2,2)$. If we put back the centroid $\epsilon$ and the width $\sigma$ in $f_{qN}$, then $S(q)$ changes to
$$
S(q:\epsilon,\sigma) = \l(\epsilon -\dis\f{2\sigma}{\sqrt{1-q}}\;,\;\epsilon +\dis\f{2\sigma}{\sqrt{1-q}}\r)\;.
$$ 
As shown in \cite{Ismail}, the even order reduced central moments of $f_{qN}$ are
\be
\barr{rcl}
\mu_{2n}(q) & = & \dis\int^{2/\sqrt{1-q}}_{-2/\sqrt{1-q}} x^{2n}\,f_{qN}(x|q)\,dx \\
&  = & (1-q)^{-n}\,\dis\sum_{r=-n}^{r=n} {2n \choose n+r} (-1)^r\,q^{r(r-1)/2}\;.
\earr \label{eq.sdm6}
\ee
Then, for example $\mu_4$, $\mu_6$ and $\mu_8$ are
\be
\barr{l}
\mu_4(q) = 2+q\;,\\
\mu_6(q) = 5+6q+3q^2+q^3\;,\\
\mu_8(q) = 14+28q+28q^2+20q^3+10q^4+4q^5+q^6\;.
\earr \label{eq.sdm7}
\ee
The $q$-Hermite polynomials $He_n(x|q)$ that are orthogonal with $f_{qN}$ as the weight function are defined by the recursion relation
\be
x\,He_n(x|q) = He_{n+1}(x|q) + \l[n\r]_q\,He_{n-1}(x|q)
\label{eq.sdm8}
\ee
with $He_0(x|q)=1$ and $He_{-1}(x|q)=0$. Note that for $q=1$, the $q$-Hermite
polynomials reduce to normal Hermite polynomials (related to Gaussian) and for
$q=0$ they will reduce to Chebyshev polynomials (related to semi-circle). The polynomials up to order 6 for example are,
\be
\barr{rcl}
He_0(x|q) & = & 1\;,\\
He_1(x|q) & = & x\;,\\
He_2(x|q) & = & x^2-1\;,\\
He_3(x|q) & = & x^3-(2+q)x\;,\\
He_4(x|q) & = & x^4-(3+2q+q^2)x^2+(1+q+q^2)\;,\\
He_5(x|q) & = & x^5 -(4+3q+2q^2+q^3)x^3 \\
& + & \l[1+q+q^2 +(2+q)(1+q+q^2+q^3)\r]x \;,\\
He_6(x|q) & = & x^6 -(5+4q+3q^2+2q^3+q^4)x^4 \\
& + & (6+9q+10q^2+9q^3+7q^4+3q^5+q^6)x^2 \\
& - & (1+q+q^2)(1+q+q^2+q^3+q^4) \;.
\earr \label{eq.sdm9}
\ee
Orthogonal property of $He_n(x|q)$'s is
\be
\dis\int^{2/\sqrt{1-q}}_{-2/\sqrt{1-q}} He_n(x|q)\,He_m(x|q)\,f_{qN}(x|q)\,dx = \l[n\r]_q!\,\delta_{mn}\;.
\label{eq.sdm10}
\ee
Using Eq. (\ref{eq.sdm10}), it is easy to derive formulas for the lower order moments as given in Eq. (\ref{eq.sdm7}). It is important to add that the $f_{qN}(x)$ defined by Eqs. (\ref{eq.sdm4}) and (\ref{eq.sdm5}) is different from the $q$-Gaussian used in Refs. \cite{qG1,qG2} and elsewhere.

\subsection{Bivariate $q$-normal distribution and Mehler expansion}

Bivariate $q$-normal distribution $f_{biv-qN}(x,y|\xi , q)$, normalized to unity as
given in \cite{Sza-1}, with $x$ and $y$ standardized variables
and defined over $S(q)$ in both $x$ and $y$ spaces, is given by
\be
\barr{l}
f_{biv-qN}(x,y|\xi , q) = f_{qN}(x|q) f_{qN}(y|q) h(x,y|\xi , q)\;;\\
\\
h(x,y|\xi ,q) \\
= \dis\prod_{k^\pr=0}^\infty \dis\frac{1-\xi^2 q^{k^\pr}}{
(1-\xi^2 q^{2k^\pr})^2 -(1-q)\,\xi\, q^{k^\pr}\,(1+\xi^2 q^{2k^\pr})\,xy +
(1-q)\xi^2 q^{2k^\pr} (x^2 +y^2)}\;,
\earr \label{eq.sdm11}
\ee
where $\xi$ is the bivariate correlation coefficient. Note that
$f_{qN}(x|q)$ and $f_{qN}(y|q)$ are the marginal densities of
$f_{biv-qN}$ and we will discuss the conditional distribution in Section \ref{cond-q}. Bivariate reduced central moments $$\mu_{rs}(q) = \int_{S(q)} x^ry^s f_{biv-qN}(x,y|\xi , q)dxdy$$ are symmetrical, i.e. $\mu_{rs}=\mu_{sr}$
and $\mu_{rs}=0$ for $r+s$ odd. Also $\mu_{20}=\mu_{02}=1$ and $\xi=\mu_{11}$. Lower order bivariate moments $\mu_{rs}$ with $r+s=4$ and $6$ are,
\be
\barr{l}
\mu_{40}(q)=\mu_{04}(q)=2+q\;,\\
\mu_{31}(q)=\mu_{13}(q)= \xi\,(2+q)\;,\\
\mu_{22}(q) = 1 + \xi^2\,(1+q)\;,\\
\mu_{60}(q)=\mu_{06}(q)=5+6q+3q^2+q^3\;,\\
\mu_{51}(q)=\mu_{15}(q)=\xi\,\mu_{60}(q)\;,\\
\mu_{42}(q)=\mu_{24}(q)=(2+q) +\xi^2\,(3+5q+3q^2+q^3)\;,\\
\mu_{33}(q)=\xi\;(2+q)^2 + \xi^3\,(1+q)(1+q+q^2)\;.
\earr \label{eq.sdm12}
\ee
These results can be derived easily from the important identity, i.e. the Poisson-Mehler formula \cite{Sza-1,Mehl,Sza-3}
\be
f_{biv-qN}(x,y|\xi , q) = f_{qN}(x|q) f_{qN}(y|q) \l[\dis\sum_{n=0}^{\infty} \dis\f{\xi^n}{\l[ n\r]_q!}\,He_n(x|q) He_n(y|q)\r]
\label{eq.sdm13}
\ee
Putting $\xi=1$ on both sides and operating over $S(q)$ gives,
\be
\delta(x-y) = f_{qN}(x|q) \l[\dis\sum_{n=0}^{\infty} \dis\f{1}{\l[n\r]_q!} He_n(x|q) He_n(y|q)\r] \;.
\label{eq.sdm14}
\ee
This formula plays an important role in SDM.

\subsection{Conditional $q$-normal distribution}
\label{cond-q}

Given the bivariate $q$-normal $f_{biv-qN}$, the conditional $q$-normal densities ($f_{CqN}$) are easily seen to be
\be
\barr{l}
f_{CqN}(x|y; \xi , q)=f_{qN}(x|q) h(x,y|\xi ,q)\;,\\
f_{CqN}(y|x; \xi , q)=f_{qN}(y|q) h(x,y|\xi ,q)\;
\earr \label{eq.sdm15}
\ee
with $h(x,y|\xi ,q)$ defined in Eq. (\ref{eq.sdm11}). A very important property of $f_{CqN}$ that follows from Eq. (\ref{eq.sdm13}) is
\be
\dis\int_{S(q)} He_n(x|q) f_{CqN}(x|y; \xi ,q) dx = \xi^n He_n(y|q)\;.
\label{eq.sdm16}
\ee
Putting $n=0$ in Eq. (\ref{eq.sdm16}), it is easy to infer that $f_{CqN}$ and hence $f_{biv-qN}$ are normalized to unity over $S(q)$. It is important to recognize that the centroid $\epsilon(y:\xi ,q)$ of $f_{CqN}(x|y;\xi ,q)$ is not zero and its variance $\sigma^2(y:\xi ,q)$ is not unity. Using Eq. (\ref{eq.sdm16}) we have
\be
\epsilon(y:\xi ,q) = \xi\,y\;,\;\;\;\sigma^2(y:\xi ,q) = (1-\xi^2)
\label{eq.sdm17}
\ee
with the centroid linear in $y$ and variance independent of $y$. In addition, we have the following formulas for the skewness and excess \cite{qMK-3,Sza-2},
\be
\barr{rcl}
\gamma_1(y:\xi ,q) & = & -(1-q)\dis\frac{\xi y}{\dis\sqrt{1-\xi^2}}\;,\\
\gamma_2(y:\xi ,q) & = & (q-1) +
(1-q)^2 \l[\dis\frac{\xi y}{\dis\sqrt{1-\xi^2}}\r]^2 + (1-q^2)\dis\frac{\xi^2}{(1-\xi^2)}\;.
\earr \label{eq.sdm18}
\ee
Thus, the $\gamma_1$ and $\gamma_2$ are zero only when $q=1$ (then the conditional distribution is a Gaussian and this result is well known \cite{Kendall}). Let us mention that the Al-Salam-Chihara
polynomials $P_n(x|y,\xi, q)$ are orthogonal with $f_{CqN}$ as the weight function giving 
\be
\dis\int_{S(q)} P_n(x|y,\xi ,q) P_m(x|y,\xi , q)f_{CqN}(x|y; \xi ,q) dx 
= \l(\xi^2\r)_n \l[n\r]_q!\, \delta_{mn}\;.
\label{eq.sdm19}
\ee
where $(\xi^2)_n=\dis\prod_{i=0}^{n-1} (1-\xi^2 q^i)$ with
$(\xi^2)_0=1$. The polynomials $P_n(x|y,\xi, q)$ satisfy the relation,
\be
P_{n+1}(x|y,\xi, q)=\l(x-\xi y q^n\r)P_n(x|y,\xi , q) -\l(1-\xi^2q^{n-1}\r)\l[n\r]_q P_{n-1}(x|y,\xi ,q)
\label{eq.sdm20}
\ee
with $P_{-1}(x|y,\xi ,q)=0$ and $P_0(x|y,\xi ,q)=1$.

Given the various properties of $q$-normal distributions and the related moments and polynomials (see \cite{Sza-1,Sza-2,Sza-3,Ismail,Mehl,Sza-4} for further details),
we will now give the formulation of SDM (statistical nuclear spectroscopy) using $q$-normal distributions and $q$-Hermite polynomials. As already mentioned, the EE basis for using these is established in \cite{qMK-1,qMK-2,qMK-3,chv1}. 

\section{Level densities with $q$-normal distribution}
\label{sec3}

Nuclear level densities are by definition statistical quantities and they are important in nuclear physics as they are measurable in some situations. More importantly, they are needed for many reaction cross section calculations and in particular for Astrophysical reaction rates calculations. See for example \cite{Zel5,app-a,app-b,SINP} and references therein.

\subsection{State density}
\label{sec31}

\begin{figure}
     \centering
         \includegraphics[width=0.9\textwidth,height=0.85\textwidth]{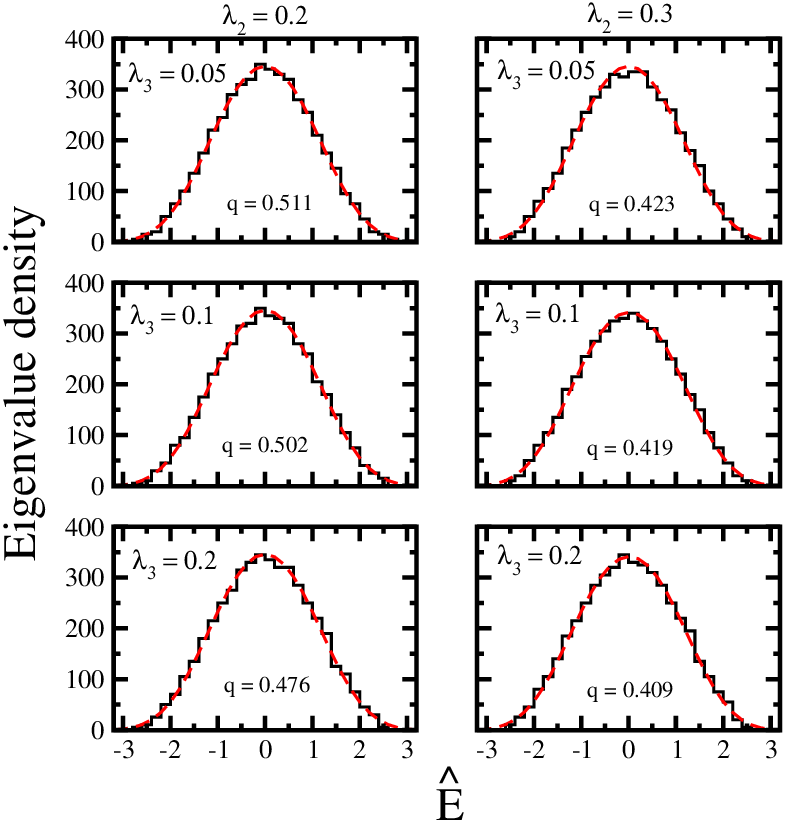}
         \caption{Eigenvalue densities for a 1000 member EGOE$(1+2+3)$ ensemble with $H = h(1) + \lambda_2 V(2) +\lambda_3 V(3)$, where $\lambda_2$ and $\lambda_3$ are interaction strengths for two-body and three-body interactions respectively. We have chosen system configuration with $N = 12$ sp states and $m = 6$ fermions with $\lambda_2 = 0.2, \, 0.3$ and $\lambda_3 = 0.05, \, 0.1, \, 0.2$. Numerical results are shown as histograms and the analytical dashed curves are obtained using Eq. \eqref{eq.sdm4} with $q$ values as given in the figure.}
        \label{fig:f1}
\end{figure}

Nuclear Hamiltonians consist of a mean-field one-body part $h(1)$, a residual two-body part $V(2)$ and a small $3$-body part (perhaps also a four-body part) 
\cite{Mos-three,three,four,t31,t32,t33,t34,t35,t36}. Thus, $H$ is $H=h+V$ with $h=h(1)$ and $V=V(2)$ or $V(2)+V(3)$ or $V(2)+V(3)+V(4)$. As we are concerned with the general formulation for nuclear level densities, explicit form of $h$ and $V$ is not essential in this paper. Say the $H$ operates in a shell model space with $m_p$ number of protons distributed in shell model single particle (sp) orbits ($j^p_1,  j^p_2, \ldots, j^p_r$) and similarly with $m_n$ number of neutrons in sp orbits
($j^n_1, j^n_2, \ldots, j^n_s$). With $N_p=\sum_i (2j^p_i+1)$ and $N_n=\sum_i (2j^n_i+1)$, total $H$ matrix dimension is $d(m_p,m_n) = \binom{N_p}{m_p} \binom{N_n}{m_n}$. Given this, we can decompose the total Hilbert space into proton configurations $\tmp=[m_p^1, m_p^2, \ldots, m_p^r]$ where $m_p^i$ is number of protons in the
orbit $j_i^p$ with $\sum_{i=1}^r\,m_p^i=m_p$ and similarly, neutron configurations $\tmn=[m_n^1, m_n^2, \ldots, m_n^s]$ where $m_n^i$ is number of neutrons in the orbit $j_i^n$ with $\sum_{i=1}^s\,m_n^i=m_n$. With these, $(\tmp, \tmn)$'s denote proton-neutron configurations. Then, the total state (or eigenvalue) density $I^H(E)$,  with $\lan \lan -- \ran \ran$ denoting trace and $\lan --\ran$ denoting average, can be written as a sum of the partial densities defined over $(\tmp, \tmn)$, 
\be
\barr{l}
I^{(m_p, m_n)}(E) = \lan \lan \delta(H-E) \ran \ran^{(m_p, m_n)} = \dis\sum_\beta\,\lan (m_p,m_n) \beta \mid \delta(H-E) \mid (m_p,m_n) \beta \ran \\
=\dis\sum_{(\tmp, \tmn)}\, \lan \lan \delta(H-E) \ran \ran^{(\tmp, \tmn)} \\
= \dis\sum_{(\tmp, \tmn)}\, d(\tmp,\tmn)\,
\rho^{(\tmp, \tmn)}(E) = \dis\sum_{(\tmp, \tmn)}\,I^{(\tmp, \tmn)}(E) \;.
\earr \label{eq.sdm21}
\ee
Eq. (\ref{eq.sdm21}) is exact and here, $d(\tmp,\tmn)$ is the dimension of the configuration $(\tmp, \tmn)$ and $\rho^{(\tmp, \tmn)}(E)$ is partial density normalized to unity. Similarly, $I^{(\tmp, \tmn)}(E)$ is the partial density normalized to $d(\tmp,\tmn)$ and $I^{(m_p, m_n)}(E)$ is total eigenvalue density normalized to the total dimension $d(m_p,m_n)$.  Without loss of generality, from now on, we will denote $(\tmp , \tmn)$ by $\wm$. The moments $M_p(\wm)$ of $\rho^{(\wm)}(E)$ are given by
\be
M_p(\wm) = \lan H^p\ran^{\wm}
\label{eq.sdm22}
\ee
with the centroid $\epsilon(\wm)=\lan H\ran^{\wm}$ and the variance $\sigma^2(\wm) =\lan H^2\ran^{\wm} -[\epsilon(\wm)]^2$.
Now, applying the result that EGOE($k$) generates $q$-normal form for the eigenvalue densities to the partial densities (this is similar to the application Gaussian form to partial densities in the past in SDM applications \cite{jbf3,KH,Zel5}), it is possible to approximate $\rho^{(\wm)}(E)$ by $\rho^{(\wm)}_{qN}(E|q)$ giving,
\be
\barr{l}
I^{(m_p,m_n)}(E) = \dis\sum_{\wm} d(\wm) \rho^{(\wm)}(E) 
\approx \dis\sum_{\wm} d(\wm) \rho^{(\wm)}_{qN}(E|q)  = \dis\sum_{\wm}
\dis\f{d(\wm)}{\sigma(\wm)}\,f_{qN}^{(\wm)}(\eh|q)\;;\\
\eh=(E-\epsilon(\wm))/\sigma(\wm)
\earr \label{eq.sdm23}
\ee
with $f_{qN}$ defined for a given $(\wm)$ and it is given by Eq. (\ref{eq.sdm4}). In addition, $\rho^{(\wm)}_{qN}(E|q)$ is defined over the interval
$$
\l(\epsilon(\wm)-\dis\frac{2\,\sigma(\wm)}{\dis\sqrt{1-q}}\;,\;\epsilon(\wm)+\dis\frac{2\,\sigma(\wm)}{\dis\sqrt{1-q}}\r)\;.
$$
The $q$ value in Eq. (\ref{eq.sdm23}) is assumed to be independent of $\wm$. 

In practice, we can either use the EGOE formulas given in \cite{qMK-1,qMK-3,KM-strn} for $q$ (formula for $q$ follows from the formula for $\gamma_2$ or $\mu_4$) or one can use $q$ as a free parameter. Formula for $q$ with EGOE($k$) representing $H$ is \cite{qMK-1},
\be
\barr{l}
q(N,m,k) = \dis\binom{N}{m}^{-1} \dis\sum_{\nu=0}^{min(k,m-k)}\;
\dis\frac{\Lambda^\nu(N,m,m-k)\;\Lambda^\nu(N,m,k)\;d(N:\nu)}{
\l[\Lambda^0(N,m,k)\r]^2} \;;\\ 
\\
\Lambda^\nu(N,m,r) =  \dis\binom{m-\nu}{r}\;\dis\binom{N-m+r-\nu}{r}\;,\\ \\
d(N:\nu)  = \dis\binom{N}{\nu}^2-\dis\binom{N}{\nu-1}^2\;.
\earr \label{eq.qform1}
\ee
Note that we are considering $m$ fermions in $N$ sp states with $H$ being a $k$-body operator. If $H$ is a (1+2)-body operators an average with appropriate weights for $q$ from Eq. (\ref{eq.qform1}) for $k=1$ and $k=2$ may suffice. Similarly, the $q$ value can be determined for
$(1+2+3)$ or $(1+2+3+4)$ body $H$. For proton-neutron systems, formulas for the second and fourth moments over $(m_p,m_n)$ spaces, derived in \cite{KM-strn} using EGOE($k$), will give the formula for $q$. In this situation, a $k$-body Hamiltonian will be of the form $H=\sum_{i+j=k} H(i,j)$ where $i$ is the body rank in proton space (space $\#1$) and $j$ is the body rank in neutron space (space $\#2$). For example, for $k=2$, $H=H_{pp}+H_{nn}+H_{pn}=H(2,0)+H(0,2)+H(1,1)$ and we put $m_1=m_p$ and $m_2=m_n$. Similarly, we put $N_1=N_p$ and $N_2=N_n$. Then assuming $H(i,j)$ are represented by independent EGOE's with matrix elements variance $v^2_H(i,j)$ in the defining spaces, we have
{\small
\be
\barr{l}
q(N_1,m_1,N_2,m_2,k)
= \dis\f{\dis\sum_{i+j=k} \,\dis\sum_{i^\pr + j^\pr =k} \;v^2_H(i,j)\, v^2_H(i^\pr , 
j^\pr)\;X(N_1, m_1 , i, i^\pr)\,Y(N_2, m_2 , j, j^\pr)}{\l[\dis\sum_{i+j=k}\,v^2_H(i,j)\;
\Lambda^0(N_1, m_1, i) \,\Lambda^0(N_2, m_2, j) \r]^2} \;; \\
\\
X(N_1, m_1 , i, i^\pr)= {\dis\binom{N_1}{m_1}}^{-1} \; \dis\sum_{\nu_1=0}^{min(i, m_1-i^\pr)}\,\Lambda^{\nu_1}(N_1, m_1, m_1-i)\, \Lambda^{\nu_1}(N_1, m_1, i^\pr)\,d(N_1:\nu_1)\;, \\
\\
Y(N_2, m_2 , j, j^\pr)= \dis{\binom{N_2}{m_2}}^{-1} \dis\sum_{\nu_2=0}^{min(j, m_2-j^\pr)}\,\Lambda^{\nu_2}(N_2, m_2, m_2-j)\, \Lambda^{\nu_2}(N_2, m_2, j^\pr)\,d(N_2 : \nu_2)\;.
\earr \label{eq.qform2}
\ee
}
This formula will be independent of the $v$ parameters if we assume that the $v^2_H(i,j)$ do not depend on $(i,j)$. 
In future, it is important to incorporate $\wm$ or $h$ and $V$ dependence of $q$. From Ref. \cite{qMK-3}, it follows that the $q$ in $\rho_{qN}^{\wm}(E|q)$ depends on $\lan hVhV \ran$ rather than on $\lan H^4\ran$. This needs to be understood better using nuclear interactions.

\begin{figure}
     \centering
         \includegraphics[width=\textwidth]{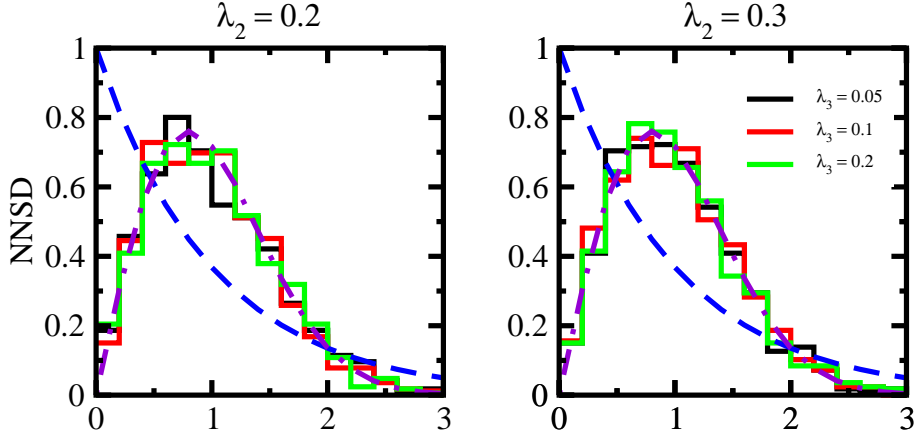}
         \caption{Spacing distribution for a 1000 member EGOE$(1+2+3)$ ensemble with $H = h(1) + \lambda_2 V(2) +\lambda_3 V(3)$, where $\lambda_2$ and $\lambda_3$ are interaction strengths for two-body and three-body interactions respectively. We have chosen system configuration with $N = 12$ sp states and $m = 6$ fermions with $\lambda_2 = 0.2, \, 0.3$ and $\lambda_3 = 0.05, \, 0.1, \, 0.2$. Eigenvalue density defined by Eq. \eqref{eq.sdm4} is employed as unfolding function using the $q$ values given in Fig. \ref{fig:f1}. Numerical results are shown as histograms. The blue dashed curve is Poisson limit and purple dot-dashed curve is the GOE limit.}
        \label{fig:f2}
\end{figure}

With $\rho_{qN}^{(\wm)}(E|q)$ in Eq. (\ref{eq.sdm23}) being a strength function \cite{Ko-01}, its centroid and width must be essentially same as in Eq. (\ref{eq.sdm17}). In order to establish this, we will use the results $\epsilon_h(\wm)=\lan h\ran^{\wm}$, $E_c(m)=\lan H\ran^m = \lan h\ran^m + \lan V\ran^m = \epsilon_h(m)+\lan V\ran^m$, $\sigma^2_H(m) \sim \sigma^2_h(m)+\sigma^2_V(m)$, $\xi=\sigma_h(m)/\sigma_H(m)$, $\sqrt{1-\xi^2}=\sigma_V(m)/\sigma_H(m)$, $\xi/\sqrt{1-\xi^2} =\sigma_h(m)/\sigma_V(m)$, 
$\lan V\ran^m \sim \lan V\ran^{\wm}$ and $\sigma^2(\wm) \sim \sigma_V(m)$. These various approximations are discussed
in the past in detail in \cite{KS}. Now, it is important to recognize that in Section \ref{cond-q}, $x=(E-E_c(m))/\sigma_H$ and $y=[\epsilon_h(\wm)-\epsilon_h(m)]/\sigma_h(m) \sim [\epsilon(\wm)-E_c(m)]/\sigma_h(m)$. In addition, $y$ is a continuous variable in Section \ref{cond-q} while we are using discrete representation in Eq. (\ref{eq.sdm21}).  The standardized variable for the strength functions in Section \ref{cond-q}
with centroid and variance given by Eq. (\ref{eq.sdm17}) is
$(x-\xi y)/\sqrt{1-\xi^2}$. Then, with the approximations given above, it is easy to see that $(x-\xi y)/\sqrt{1-\xi^2} \sim (E-\epsilon(\wm))/\sigma(\wm)$ and this gives correctly the $\eh$ in $f^{(\wm)}_{qN}$. With this, we can use the formulas for $\gamma_1$ and $\gamma_2$ given by Eq. (\ref{eq.sdm18}) to add corrections to $f^{(\wm)}_{qN}(\eh |q)$. To this end, we will consider a Gram-Charlier (GC) like expansion \cite{Kendall} of $\rho^{(\wm)}(E)$ in terms of $q$-Hermite polynomials in $(\wm)$ space. Then, as a function of the standard variable $\eh$, we have,
\be
\barr{l}
\eta^{(\wm)}(\eh) = f^{(\wm)}_{qN}(\eh |q)\l[1+\dis\sum_{n=1}^{\infty} \dis\f{c_n}{[n]_q!}\;He_n(\eh|q)\r]\;\\
c_n=\dis\int_{S(q)} He_n(\eh|q)\, \eta^{(\wm)}(\eh)\,d\eh\;.
\earr \label{eq.sdm24}
\ee
Note that $\rho(E)\,dE = \eta(\eh)\,d\eh$.
Formula for the expansion coefficients $c_n$ follow from Eq. (\ref{eq.sdm10}). It is easy to see that $c_1=0$, $c_2=0$, $c_3=\gamma_1$ and $c_4=\gamma_2+1-q$ where $\gamma_2$ is the value of $\gamma_2$ for $\eta^{(\wm)}(\eh)$. Used here are the formulas for $q$-Hermite polynomials given in Eq. (\ref{eq.sdm9}). If we chose the $q$ such that the $\gamma_2$ of $\eta^{(\wm)}(\eh)$ and $f^{(\wm)}_{qN}(\eh |q)$ are same, then $c_4=0$.
The GC expansion given by Eq. (\ref{eq.sdm24}), but in $m$-particle spaces without $\wm$ decomposition, is used recently in \cite{Verb4,chv2} for the so called unfolding of the spectrum. As our interest is in the smoothed forms, 
we can use the approximation that truncating the expansion in Eq. 
(\ref{eq.sdm24}) to $n \leq 4$ is good and this gives
\be
\barr{l}
\eta^{(\wm)}(\eh) = f^{(\wm)}_{qN}(\eh |q) \l[1+\dis\frac{\gamma_1(\wm)}{[3]_q!} He_3(\wm|q) +\dis\f{(\gamma_2(\wm)+1-q)}{[4]_q!} He_4(\wm|q)\r]\;;\\
\\
\gamma_1(\wm) \approx -(1-q)\l[\dis\f{\epsilon(\wm)-E_c(m)}{\sigma(\wm)}\r]\;,\\
\\
\gamma_2(\wm) = (q-1) + (1-q)^2 \l[\dis\f{\epsilon(\wm)-E_c(m)}{\sigma_{\wm}}\r]^2 + (1-q^2) \dis\f{\sigma^2_h(m)}{\sigma^2_V(m)}\;.
\earr \label{eq.sdm25}
\ee
Here, the formulas for $\gamma_1$ and $\gamma_2$ are obtained using Eq. (\ref{eq.sdm18}) and the approximations mentioned above. Eqs. (\ref{eq.sdm23}) and (\ref{eq.sdm25}) will allow us to calculate state densities generated by $H$ using $q$-normal distribution.

Let us emphasize that, unlike a Gaussian, the $f^{(\wm)}_{qN}(\eh)$ are defined over the range $-2/\sqrt{1-q} \le \eh \le +2/\sqrt{1-q}$ with $q$ given by an average $\gamma_2$ value (assumed to be independent of $\wm$). Given $\gamma_2$, we have the result $q=1+\gamma_2$ from Section \ref{q-nrm}. Then for example, for $\gamma_2=-0.2$ the range is $-4.47 \le \eh \le 4.47$ , for $\gamma_2=-0.3$ the range is $-3.65 \le \eh \le 3.65$ and for $\gamma_2=-0.4$ the range is $-3.16 \le \eh \le 3.16$. A typical value is $\gamma_2 \sim -0.3$ (see \cite{KM-strn} for examples).

As examples, we show eigenvalue densities for a 1000 member EGOE$(1+2+3)$ ensemble with $H = h(1) + \lambda_2 V(2) +\lambda_3 V(3)$, where $\lambda_2$ and $\lambda_3$ are interaction strengths for two-body and three-body interactions respectively. We have chosen system configuration with $N = 12$ sp states and $m = 6$ fermions with $\lambda_2 = 0.2, \, 0.3$ and $\lambda_3 = 0.05, \, 0.1, \, 0.2$. We choose $V(2)$ and $V(3)$ to be independent EGOE's and $h(1)$ is defined by fixed sp energies $\epsilon_i = i + 1/i$; $i = 1, \, 2,\, \cdots,\, N$. Numerical results are shown as histograms and the analytical dashed curves are obtained using Eq. \eqref{eq.sdm4}.  Agreement between theory and numerics is excellent.

Unfolding the eigenvalue spectrum using the smoothed state density given by $f_{qN}$ (Eq. \eqref{eq.sdm4}), we have obtained the nearest neighbor spacings distributions for a 1000 member EGOE$(1+2+3)$ ensemble with $H = h(1) + \lambda_2 V(2) +\lambda_3 V(3)$, where $\lambda_2$ and $\lambda_3$ are interaction strengths for two-body and three-body interactions respectively. The choice of the parameters is same as in Fig. \ref{fig:f1}. Results are shown in Fig. \ref{fig:f2}. The blue dashed curve is Poisson limit and purple dot-dashed curve is the GOE limit. As can be seen from this figure, for the choice of parameters made, the spacing distribution follows GOE and thus, we are in the many-body chaotic regime.
 
\subsection{$J$-decomposition and level densities}
\label{sec32}

In order to obtain level densities (these can be compared with experimental data), we need to carry out $J$ decomposition of state densities. Level density $I^{(m_p,m_n),J_r}(E)$ corresponds to number of levels with the given angular momentum $J_r$ at energy $E$ in unit energy interval. Note that each ($\wm$) carries definite parity and therefore parity decomposition of state and level densities is direct with the use of Eq. (\ref{eq.sdm23}). Similarly, with fixed-$(m_p,m_n)$ densities, for isospin $T$ invariant $H$'s, it is easy to obtain isospin decomposition of state and level densities if needed. Let us consider the $J$ decomposition. 

A simpler approach is to employ Bethe's spin-cutoff factor \cite{Bethe1,Bethe2} with energy dependence. Using energy dependent spin-cutoff factors $\sigma_J(E)$, we have for the level density 
\be
\barr{l}
I^{(m_p,m_n),J_r}(E) = C_{J_r}(E)\;I^{(m_p,m_n)}(E)\;, \\
C_{J_r}(E)=\dis\frac{(2J_r+1)}{\dis\sqrt{8\pi}\; \sigma^3_J(E)}  \exp
-(2J_r+1)^2/8\sigma_J^2(E)\;;\;\;\sigma_J^2(E) = \lan J_Z^2\ran^E\;.
\earr \label{eq.sdm26}
\ee
This is employed in many SDM studies \cite{jbf3,KH}. The $J_z^2$ expectation value can be calculated using 
\be
\lan J_Z^2\ran^E = \dis\f{\lan J_Z^2 \delta(H-E)\ran^m}{\lan \delta(H-E)\ran^m} = I^m_{J_Z^2}(E)/I^m(E)\;.
\label{eq.sdm27}
\ee 
The spin-cutoff densities $I^m_{J_Z^2}(E)$ can be approximated by $q$-normal distribution with or without decomposing it into partial densities and similarly the state density $I^m(E)$. Similar approach using Gaussian densities was used in the past \cite{KH}. It is also possible to add $q$-Hermite polynomial corrections (see Section \ref{sec4}).

Alternatively, angular momentum decomposition can be directly carried out by constructing fixed-$J$ densities using 
\be
I^{(m_p, m_n),J}(E) = \dis\sum_{(\tmp, \tmn) \in J}\;
I^{(\tmp, \tmn),J}_{qN}(E|q)\;.
\label{eq.sdm28}
\ee
It is possible to evaluate exact values of the centroid energies
$E_c((\tmp,\tmn),J) = \lan H \ran^{(\tmp, \tmn),J}$ and variances
$\sigma^2((\tmp,\tmn),J) = \lan H^2  \ran^{(\tmp, \tmn),J} -
[E_c((\tmp,\tmn),J)]^2$ \cite{jbf4,Wong,KH,Zel3,HGZ}, though computationally extensive, and then construct 
fixed-$J$ $q$-normal partial densities $I^{(\tmp, \tmn),J}_{qN}(E|q)$ in Eq. (\ref{eq.sdm28}). Again, $q$ may be used as
a free parameter and perhaps it is good to use the parametrization $q=q_0+q_1 J(J+1)$ with $q_0$ and $q_1$ free parameters. A better approach is to develop a theory for the $J$ and $(\wm)$ dependence of $q$ but this is not yet available.
 
\subsection{Ground state energy}
\label{sec33}

Besides the $J$ decomposition, it is necessary to determine the ground state energy $E_g$ (with reference to this excitation energies are defined). This is needed though the $q$-normal distributions are bounded. One approach is to use the so-called Ratcliff procedure. Say in experimental data of a nucleus
all level with $J^\pi$ assignments are known up to an excitation energy $E_R$ (i.e. complete spectra is known up to and including $E_R$, the reference energy) and number of states up to $E_R$ is say $N_R$ with the $J$ value for the last
level is $J_R$. Note that $N_R=\sum (2J+1)$ with the sum over all levels up to $J_R$. Now, inverting the following equation will give $E_g$,
\be
N_R -(2J_R+1)/2 = \dis\int_{Q}^{E_g+E_R}\;I^{(m_p,m_n)}(E) \;dE\;.
\label{eq.sdm29}
\ee
With $I^{(m_p,m_n)}(E)$ decomposed into partial densities as in Eq. (\ref{eq.sdm23}), $Q$ is the lowest of $\epsilon(\wm)-[2\sigma(\wm)/(1-q)]$; see Section \ref{sec31}. Instead of $I^{(m_p,m_n)}(E)$, if we use fixed-$J$ densities $I^{(m_p,m_n),J}(E)$,
then suitable modification of Eq. (\ref{eq.sdm29}) is needed.
This method of Ratcliff will work best when $E_R$ is sufficiently large.  This is adopted in many papers in the past \cite{KH} and more recently in \cite{Kar} and in \cite{KoChv}. A better method, if possible, is to use the so-called exponential convergence method \cite{HVZ}. This is based on the fact that in the process of successive truncation to dimension $d$ in the diagonalization of the shell model
$H$ matrix of dimension $D$, the ground state energy $E_g$ converges exponentially, $E[d]=E[d \rightarrow D] + C_0 \exp -\gamma d$ where $C_0$ is some constant. This is used in many recent level density studies using SDM \cite{Zel3,Zel4,Zel5}. 
 
\section{Shell model orbit occupancies using $q$-normal and $q$-Hermite polynomials}
\label{sec4}

\begin{figure}
     \centering
         \includegraphics[width=0.9\textwidth,height=0.85\textwidth]{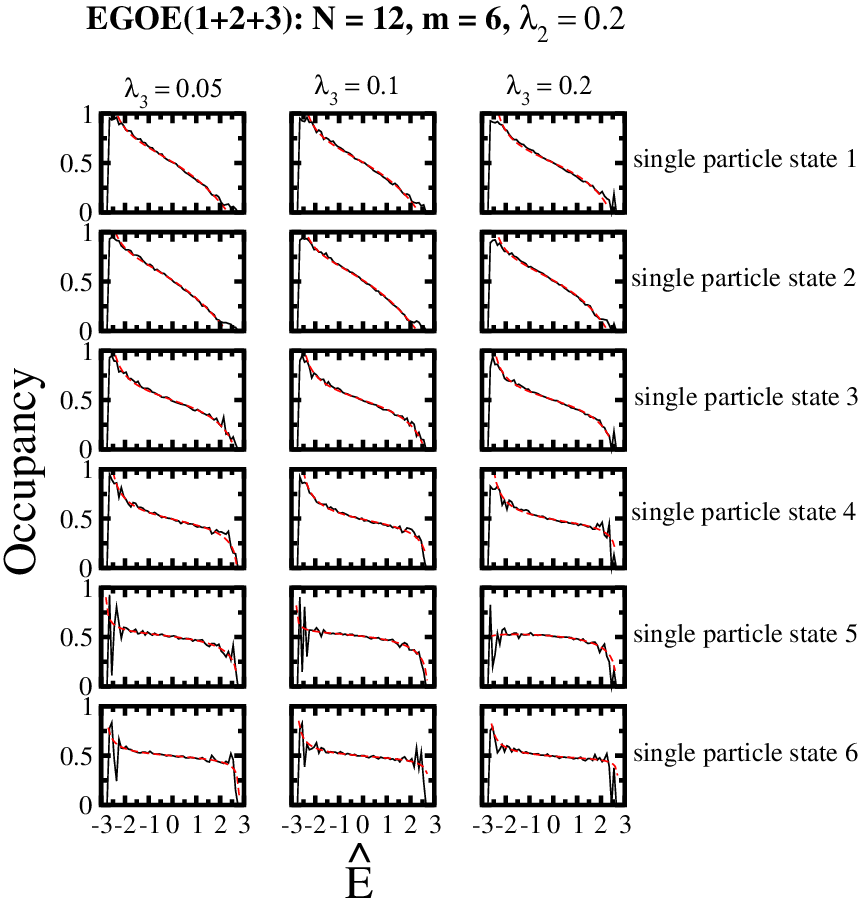}
         \caption{Single particle occupancies for first six levels for a 1000 member EGOE$(1+2+3)$ ensemble defined by $H = h(1) + \lambda_2 V(2) + \lambda_3 V(3)$ with $N = 12$, $m = 6$, $\lambda_2 = 0.2$ and $\lambda_3 = 0.05, \, 0.1, \, 0.2$. Numerical results (black curves) are compared with analytical curves (red dashed) obtained using Eq. \eqref{eq.sdm31}.}
        \label{fig:f3}
\end{figure}

Shell model orbit occupancies are measurable and for example in the last decade there are several experiments by Schiffer and collaborators measuring proton and neutron orbit occupancies in nuclei that are candidates for neutrinoless double beta decay; see \cite{KoChv,KoAps} and references therein. Similarly, they are also needed in many applications, see for example \cite{app-ma,app6}. Occupancies are expectation values of the orbit number operators. Then, for a shell model orbit $\alpha$, occupancy is given by $\lan n^x_\alpha\ran^E$ where $n_\alpha$ is the number operator for the orbit $\alpha$ and $x$ is proton or neutron. Usually ground state occupancies are measurable but in many applications one needs orbit occupancies for excited states (see Section \ref{sec53} for an example). One approach to obtain occupancies is to use Eq. (\ref{eq.sdm14}) giving,
{\small
\be
\lan n^x_\alpha\ran^E = \dis\frac{\lan n^x_\alpha \, \delta(H-E)\ran^{(m_p,m_n)}}{\lan \delta(H-E)\ran^{(m_p,m_n)}} = \dis\f{
\dis\sum_{r=0}^{\infty} \dis\f{1}{[r]_q!}\, \lan n^x_\alpha\,He_r(\ehh |q)\ran^{(m_p,m_n)}\,He_r(\eh |q)}{1+\dis\sum_{r=1}^{\infty} \dis\f{1}{[r]_q!}\, He_r(\eh |q)}\;.
\label{eq.sdm30}
\ee}
Here, $\eh = (E-E_c)/\sigma$, $\ehh = (H-E_c)/\sigma$, $E_c=\lan H\ran^{(m_p,m_n)}$ and $\sigma^2=\lan H^2 \ran^{(m_p,m_n)}-E_c^2$.
Also, $q$ here is related to the fourth moment of the state density $I^{(m_p,m_n)}(E)$; see Sections \ref{q-nrm} and \ref{sec31}. Either one can use the EGOE formula given by Eq. (\ref{eq.qform2}) for $q$ or use it as a free parameter. Truncation of the expansion given by Eq. (\ref{eq.sdm30}) to first two terms may not be adequate and also it may not be good for obtaining occupancies near the ground state. A better approximation is to treat $[\lan n^x_\alpha\ran^{(m_p,m_n)}]^{-1}\,\lan n^x_\alpha \, \delta(H-E)\ran^{(m_p,m_n)}$ as a probability density and assume that this follows the state density and hence takes $q$-normal form. Then, $\lan n^x_\alpha\ran^E $ is a ratio of two $q$-normal distributions giving
\be
\lan n^x_\alpha\ran^E = \lan n^x_\alpha\ran^{(m_p,m_n)} \dis\f{\sigma(m_p,m_n)\,f_{n^x_\alpha\,:\,qN}^{(m_p,m_n)}(\eh^\prime |q)}{\sigma(n^x_\alpha : m_p,m_n) f^{(m_p,m_n)}_{qN}(\eh |q)}\;.
\label{eq.sdm31}
\ee
The $\eh = [E-E_c(m_p,m_n)]/\sigma(m_p,m_n)$ where the state density centroid is $E_c(m_p,m_n)=\lan H\ran^{(m_p,m_n)}$ and variance is $$\sigma^2(m_p,m_n)= \lan H^2\ran^{(m_p,m_n)}- [E_c(m_p,m_n)]^2 \,.$$ Similarly, the $n^x_\alpha$-density centroid is
$$
E_c(n^x_\alpha : m_p,m_n)= \l[\lan n^x_\alpha\ran^{(m_p,m_n)}  \r]^{-1}\,\lan n^x_\alpha \, H\ran^{(m_p,m_n)}
$$
and variance is
$$
\sigma^2(n^x_\alpha : m_p,m_n) = \l[\lan n^x_\alpha\ran^{(m_p,m_n)}  \r]^{-1}\,\lan n^x_\alpha H^2\ran^{(m_p,m_n)} - \l[E_c(n^x_\alpha : m_p,m_n)\r]^2\;. 
$$
With these $\eh^\prime$ in Eq. (\ref{eq.sdm31}) is $(E-E_c(n^x_\alpha : m_p,m_n))/\sigma(n^x_\alpha : m_p,m_n)$. In principle,
the $q$ in the state density need not be same as the $q$ in the $n^x_\alpha$ density. However, as a first step one may use the EGOE formula for $q$ as given by Eq. (\ref{eq.qform2}). The formulation given by Eq. (\ref{eq.sdm30}) with Gaussians was used for example in \cite{CF} (see also \cite{MM,Wong,KH}) and the formulation given by  Eq. (\ref{eq.sdm31}) with Gaussians was used for example in \cite{ManKC} (see also \cite{MM,KH}). As an example we show in Figure \ref{fig:f3}, single particle occupancies for first six levels for a 1000 member EGOE$(1+2+3)$ ensemble defined by $H = h(1) + \lambda_2 V(2) + \lambda_3 V(3)$ with $N = 12$, $m = 6$, $\lambda_2 = 0.2$ and $\lambda_3 = 0.05, \, 0.1, \, 0.2$. We choose $V(2)$ and $V(3)$ to be independent EGOE's and $h(1)$ is defined by fixed sp energies $\epsilon_i = i + 1/i$; $i = 1, \, 2,\, \cdots,\, N$. Numerical results (black curves) are compared with analytical curves (red dashed) obtained using Eq. \eqref{eq.sdm31}. The agreement is very good with deviations at the spectrum edges.

Though the above two methods are useful in certain situations, in general it is necessary to decompose the shell model space into proton-neutron configurations as in Section \ref{sec31}. This is quite appropriate also because the states with a fixed $(\tmp , \tmn)$ configuration are eigenstates of the number operator $n^x_\alpha$ giving $n^x_\alpha \l. \l|(\tmp ,\tmn)\beta \r.\ran = m_x^\alpha(\tmp , \tmn)\, \l. \l|(\tmp ,\tmn)\beta \r.\ran $. Therefore,
{\small
\be
\barr{l}
\lan n^x_\alpha \ran^E = 
\dis\sum_{(\tmp, \tmn)}\;\dis\f{\lan\lan n^x_\alpha \delta(H-E)\ran\ran^{(\tmp , \tmn)}}{\lan\lan \delta(H-E) \ran\ran^{(\tmp ,\tmn)}} = 
\dis\f{\dis\sum_{(\tmp, \tmn)}\;m_x^\alpha(\tmp, \tmn)\;I^{(\tmp, \tmn)}(E)}{\dis\sum_{(\tmp, \tmn)} I^{(\tmp, \tmn)}(E)}\;\\
\\
\approx \dis\f{\dis\sum_{(\tmp, \tmn)}\;m_x^\alpha(\tmp, \tmn)\;I^{(\tmp, \tmn)}_{qN}(E|q)}{\dis\sum_{(\tmp, \tmn)} I^{(\tmp, \tmn)}_{qN}(E|q)}\;\\ 
\earr\label{eq.sdm32}
\ee}
The first line of this equation is exact. In the second line,  the $q$-normal form is applied to the partial densities $I^{(\tmp, \tmn)}(E)$ and to these it is possible to add corrections using $q$-Hermite polynomials [see Eq. (\ref{eq.sdm25})] just as in Section \ref{sec31}. Now, determining say for example the ground state energy as in
Section \ref{sec33}, we will obtain ground state orbit occupancies. Similarly one can also obtain orbit occupancies for excited states. It is also possible to incorporate $J$ projection. Also, as in Section \ref{sec31}, we need to assume that the $q$ is a parameter or determine its value using  Eq. (\ref{eq.qform2}). 

\section{Transition strengths and bivariate $q$-normal distribution}
\label{sec5}

Given a transition operator $\co$ acting on an $H$ eigenstate $\l.\l| E_i\r. \ran$ of a $m$
particle system, it will in general connect to eigenstates $\l.\l| E_f \r.\ran$ of the same system or a different system depending on the nature of $\co$. Then, $|\lan E_f \mid \co \mid E_i\ran|^2$ is called  transition strength. Multiplying it with the state densities at the two energies define a bivariate distribution, called  transition strength density $I^H_\co(E_i, E_f)$,
\be
I^H_\co(E_i, E_f) = I(E_f) \l|\lan E_f \mid \co \mid E_i \ran\r|^2 I(E_i)\;.
\label{eq.sdm33}
\ee
Then, the normalized strength density is $\rho^H_\co(E_i,E_f)=
\l[\lan\lan \co^\dagger \co \ran\ran\r]^{-1}\,I^H_\co(E_i, E_f)$. 
Transition strengths and transition strength sums are in some situations measurable and more importantly they are needed (as a function of excitation energies) for many applications such as in calculating $\beta$-decay rates, in neutrinoless double $\beta$-decay transition matrix elements, in time reversal and parity breaking studies and so on \cite{FKPT,app-tst1,app-tst2,app-tst3,app-tst4,app-tst5}.
Representing $H$ by a EGOE($k$) and $\co$ by another independent EGOE($t$), it is shown recently that $I_{\co}(E_i,E_f)$ is close to a $q$-bivariate normal distribution (in all past applications bivariate Gaussian form is used \cite{FKPT,KH,KoChv}). Although this is established in \cite{qMK-2} using transition operators that are $t$-body in nature (then initial and final spaces are same), from the bivariate cumulants derived in \cite{KM-strn} for $\beta$ decay and double $\beta$ decay type operators and particle transfer operators, it can be argued that the bivariate $q$-normal form applies in general. Results in the Tables 2, 3 and 4 in \cite{KM-strn} show that the cumulants $k_{rs} \approx k_{sr}$ (cumulants are shape parameters and related in a simple manner to the moments $\mu_{rs}$ \cite{Kendall}) as needed for a symmetrical distribution like bivariate $q$-normal; see Eq. (\ref{eq.sdm12}). Note that the bivariate moments $M_{PQ}$ of $\rho^H_\co(E_i,E_f)$ are,
\be
M_{PQ} = \dis\f{\lan \co^\dagger H^P \co H^Q \ran}{\lan \co^\dagger \co \ran}\;.
\label{eq.sdm34}
\ee
Bivariate $q$-normal with  marginal centroids
$(\varepsilon_i,\varepsilon_f)$, variances $(\sigma^2_i, \sigma^2_f)$ and bivariate correlation coefficient $\xi$ is given by,
\be
\barr{l}
\rho_{biv-qN}(E_i,E_f:\varepsilon_i,\varepsilon_f,\sigma_i,\sigma_f,\zeta ; q) = \dis\f{1}{\sigma_i \sigma_f}\; f_{biv-qN}(\hat{E}_i, \hat{E}_f|\xi , q)\;;\\
\hat{E}_i =(E_i-\varepsilon_i)/\sigma_i\;,\;\;\;\hat{E}_f =(E_f-\varepsilon_f)/\sigma_f
\earr \label{eq.sdm35}
\ee
with $f_{biv-qN}$ defined by Eq. (\ref{eq.sdm11}). 
With $H$ represented by EGOE($k$) and $\co$ a $t$-body operator represented by an independent EGOE($t$), the bivariate correlation coefficient $\xi$ is given by \cite{qMK-2,KM-strn} (appropriate for identical nucleons),
\be
\xi = \dis\sum_{\nu=0}^{min(t,m-k)}\; 
\dis\frac{\Lambda^\nu(N,m,m-t)\;\Lambda^\nu(N,m,k)\;d(N : \nu)}{\binom{N}{m}\;
\Lambda^0(N,m,k)\;\Lambda^0(N,m,t)} \,.
\label{eq.formu}
\ee
Similarly, formula for $q$ follows from Eq. (\ref{eq.qform1}).
The functions $\Lambda^\nu(N,m,k)$ and $d(N : \nu)$ are defined in Eq. (\ref{eq.qform1}).

In the shell model spaces, just as with the level densities and occupancies, it is necessary to deal with proton-neutron partitioning. There are complications in  decomposing $I^H_\co(E_i, E_f) $
into configuration partial densities such that the partial strength densities are always positive definite \cite{KoAps}. Then, proceeding as suggested in \cite{FKPT}, for $H=h+V$, we have the bivariate convolution form $I^{H=h+V}_{\co}(E_i,E_f) = \int I^{h}_{\co}(x,y)\,\rho^{V}_{\co} (E_i-x,E_f-y)\,dx\,dy$. Here, $I^{h}_{\co}$ is the transition strength density generated by the one-body part $h(1)$ of $H$ and $\rho^{V}_{\co}$
is the normalized transition strength density due to interactions (as mentioned before, for nuclei $V$ is $2-$body with some small $3-$ and $4-$body parts). Using ($\wm$)'s, in many situations it is possible to construct $I^{h}_{\co}$ and, from the discussion above, $\rho^{V}_{\co}$ is a bivariate $q$-normal distribution. 
Then we have,
\be
\barr{l}
\l|\lan E_f \mid \co \mid E_i \ran\r|^2 = \dis\sum_{\wm_i,\wm_f} 
\dis\frac{
I^{\wm_i}_{qN}(E_i|q_c) I^{\wm_f}_{qN}(E_f|q_c)}{I^{m_i}(E_i|q_c) I^{m_f}(E_f|q_c)}
\l|\lan \wm_f \mid \co \mid \wm_i\ran\r|^2 \\ 
\times \dis\frac{\rho^V_{\co : biv-qN}\l(E_i , E_f\,:\, 
\ce(\wm_i), \ce(\wm_f),
\sigma(\wm_i), \sigma(\wm_f), \xi(\wm_i , \wm_f) ; q_V\r)}{
\rho_{qN}^{\wm_i}(E_i:q_c)\;
\rho_{qN}^{\wm_f}(E_f:q_c)}\;. 
\earr \label{eq.sdm36}
\ee
Here, $\l|\lan \wm_f \mid \co \mid
\wm_i  \ran\r|^2 = [d(\wm_i) d(\wm_f)]^{-1}\,\sum_{\gamma_i , \gamma_f}\l|\lan \wm_f,\gamma_f \mid \co \mid \wm_i, \gamma_i\ran\r|^2$ and for many different types of $\co$ operators, it is easy to derive formulas for this \cite{jbf3,Wong,KH,DM}. For the
marginal centroids and variances of $\rho^V_{\co : biv-qN}$ used often are the approximations (assumed to be independent of $\co$), $\ce(\wm_i) = \lan H \ran^{\wm_i}$,
$\ce(\wm_f) = \lan H \ran^{\wm_f}$, $\sigma^2(\wm_i) = \lan H^2 \ran^{\wm_i} - [\ce(\wm_i)]^2$ and $\sigma^2(\wm_f) = \lan H^2 \ran^{\wm_f} - [\ce(\wm_f)]^2$. The correlation coefficient $\xi$ depends on $\co$ and formulas for $\xi$ can be derived by assuming that it depends on $(m_p,m_n)$ of the initial and final states but not on ($\wm_i$) and ($\wm_f$). These formulas are given in the next two subsections. For $q_V$ one may use the formula given by Eq. (\ref{eq.qform2}) and $q_c$ can be treated as a free parameter or use $q_c$ as determined by level densities (see Section \ref{sec31}).

Before going further, it is important to mention that in practical
applications and in comparing with data, often we need $J$ decomposition of transition strengths given by Eq. (\ref{eq.sdm36}). An approximate method for $J$ projection is again to use energy dependent spin-cutoff factors as described for example in \cite{FKPT,KoAps}. There are additional complications as in general transition operators are tensor operators with respect to $J$ and hence it is also important to consider transition strength densities with reduced matrix elements of transition operators; see \cite{DFW} for discussion and some results related to this. Besides transition strengths, it is also important to directly obtain non-energy weighted transition strength sums given by $\lan \co^\dagger \co\ran^E$ and and also moments of the distribution of 
strengths orginating from an eigenstate with energy $E$. These are discussed in
Sections \ref{sec53} and \ref{sec54}.

\subsection{Correlation coefficient for $\beta$ and double $\beta$ decay type transition operators}
\label{sec51}

Given a state with $(m_p,m_n)$, $\beta$ and double $\beta$ decay type transition operators $\co$ generate the final states with $(m_p \pm k_0, m_n \mp k_0)$; $k_0=1$ for $\beta$ ($\beta^{\pm}$) decay and similarly $k_0=2$ for double $\beta$-decay. With $H$ a $k$-body operator, it will be of the form $H=\sum_{i+j=k} H(i,j)$ where $H$ is $i$-body in proton space and $j$-body in neutron space. In this Section, we will use $(m_1,m_2)=(m_p,m_n)$ and $(N_1,N_2)=(N_p,N_n)$. Representing $H(i,j)$ and $\co$ by appropriate
independent EGOE's with matrix elements variances $v_H^2(i,j)$ and $v_\co^2$ respectively in the defining spaces(see \cite{KM-strn} for the definition of these ensembles) both in $(m_p,m_n)$ space, formula for the bivariate correlation coefficient $\xi$ is given by \cite{KM-strn}, 
{\small
\be
\barr{l}
\xi(N_1,m_1,N_2,m_2,k,k_0) = \\
\dis\f{M_{11}(N_1,m_1,N_2,m_2,k,k_0)}{\binom{N_1-m_1}{k_0} \binom{m_2}{k_0}\;\l[M_2(N_1,m_1,N_2,m_2)\;M_2(N_1,m_1+k_0,N_2,m_2-k_0)\r]^{1/2}}\;;\\
\\
M_2(N_1,m_1,N_2,m_2)=\dis\sum_{i+j=k} v^2_H(i,j) \Lambda^0(N_1,m_1,i) \Lambda^0(N_2,m_2,j)\;,\\
M_{11}(N_1,m_1,N_2,m_2,k,k_0) = \l\{\dis\binom{N_1}{m_1}
\;\dis\binom{N_2}{m_2}\r\}^{-1} \dis\sum_{i+j=k} V^2_H(i,j) 
\dis\binom{N_1-k_0}{m_1} \dis\binom{N_2-k_0}{m_2-k_0} \\
\times \l[\dis\sum_{\nu_1=0}^i X_{11}(N_1, m_1, k_0, i, \nu_1)\r]\;
\l[\dis\sum_{\nu_2=0}^j Y_{11}(N_2, m_2, k_0, j, \nu_2)\r] \;;
\\ \\
X_{11}(N_1, m_1, k_0, i, \nu) = \l[\dis\binom{N_1}{k_0}\,d(N_1:\nu)\r]^{1/2} \;\l|U(f_{m_1+k_0}\,\overline{f_{m_1}}\,
f_{m_1+k_0}\,f_{m_1}\,;\,f_{k_0}\,\nu)\r| \\
\\ 
\times
\;\l[\Lambda^{\nu}(N_1, m_1, m_1-i)\,\Lambda^{\nu}(N_1, m_1+k_0, 
m_1+k_0-i)\r]^{1/2}\;, \\
\\
Y_{11}(N_2, m_2, k_0, j, \nu) = X_{11}(N_2, m_2-k_0, k_0, j, \nu)\;,
\\
\l| U(f_m,\, \overline{f_p},\, f_m,\, f_p\, ;\,f_{m-p}\,\nu) \r| = \l[ \dis\frac{{\binom{N+1}{\nu}}^2 \binom{m-\nu}{p-\nu} \binom{N-\nu -p}{m-p}\;
(N-2\nu +1)}{{\binom{N-m+p}{p}}^2 \binom{N}{m-p}\;(N+1)}\r]^{1/2}\;.
\earr \label{eq.qform3}
\ee}
Note that, in Young tableaux notation, $f_r = \{1^r\}$, $\overline{f_r}=\{1^{N-r}\}$ and $\nu=\{2^\nu , 1^{N-2\nu}\}$ with $N=N_1$ or $N_2$ as appropriate. It is important to mention that although the formula for $\xi$ is derived using EGUE, it applies to EGOE as shown in \cite{KM-strn}.

\subsection{Correlation coefficient for Electro-magnetic type transition operators}
\label{sec52}

Electromagnetic transition (EM) operator are one-body operators and they are sum of the operators in proton space and neutron space. Let us consider a general $k_0$-body operator in proton-neutron space. Then, $\co = \co_1(k_0) + \co_2(k_0)$ with $\#1$ and $\#2$ denoting $p$ and $n$ spaces. For EM operators, $k_0=1$.  Again, using $H=\sum_{i+j=k} H(i,j)$ and representing $H(i,j)$, $\co_1(k_0)$ and $\co_2(k_0)$ by independent EGOE ensembles with matrix element variances $v^2_H(i,j)$, $v^2_{\co_1}$ and $v^2_{\co_2}$ in the defining spaces, formula for the correlation coefficient can be written down following the procedure given in \cite{KM-strn}. The final result is,
\be
\barr{l}
\xi(N_1,m_1,N_2,m_2,k,k_0) = \dis\f{M_{11}(1)+M_{11}(2)}{M_{00} \;\cam} \;;\\
\\
M_{11}(1) = v^2_{\co_1} \dis\sum_{i+j=k} v^2_H(i,j)\,{\binom{N_1}{m_1}}^{-1} \Lambda^0(N_2,m_2,j)\, X(N_1,m_1,k_0,i)\;, \\
\\
M_{11}(2) = v^2_{\co_2} \dis\sum_{i+j=k} v^2_H(i,j)\,{\binom{N_2}{m_2}}^{-1} \Lambda^0(N_1,m_1,i)\,X(N_2,m_2,k_0,j)\;,\\
\\ 
X(N_a,m_a,k_0,p) =
\l[\dis\sum_{\nu =0}^{min(p,m_a-k_0)} \Lambda^\nu(N_a,m_a,k_0) \Lambda^\nu(N_a,m_a,m_a-p)\,d(N_a :\nu)\r]\;,\\
\\
M_{00} = v^2_{\co_1} \Lambda^0(N_1,m_1,k_0) + v^2_{\co_2} \Lambda^0(N_2,m_2,k_0)\;,\\
\\
\cam = \dis\sum_{i+j=k} v^2_H(i,j) \Lambda^0(N_1,m_1,i) \Lambda^0 (N_2,m_2,j)\;.
\earr \label{eq.qform4}
\ee
Let us recollect that $(m_1,m_2)=(m_p,m_n)$ and $(N_1,N_2)=(N_p,N_n)$. In addition to EM type and  $\beta$ and double $\beta$-decay type, it is also possible to derive a formula for $\xi$ for particle transfer operators using the results of Section 7 in \cite{KM-strn}.

\subsection{Transition strength sums} 
\label{sec53}

Strength sums (also called non-energy weighted sum rule (NEWSR) quantities) generated by one-body transition operators such as  GT operator, quadrupole ($E2$) transition operator are important as they are often measurable and carry new nuclear structure information. 
Starting with Eq. (\ref{eq.sdm36}), and integrating over the final state energies $E_f$, it is easy to obtain a formula for the transition strength sums $\lan \co^\dagger \co\ran^E$ (i.e. for the sum of the transition strengths originating from an eigenstate of $H=h+V$ with energy $E$) again involving $q$-normal distribution. Then we have,
\be
\lan \co^\dagger \co\ran^{E} = \dis\sum_{(\tmp ,\tmn)}\;\dis\f{I_{qN}^{(\tmp ,\tmn)}(E|q)}{I_{qN}^{(m_p ,m_n)}(E|q)}\;\lan \co^\dagger \co\ran^{(\tmp ,\tmn)}\;.
\label{eq.sdm37}
\ee
Formulas for $\lan \co^\dagger \co\ran^{(\tmp ,\tmn)}$ can be written down for a variety of one and two-body operators \cite{KH}. Also, the energies $E$ here follow from the procedure given in \ref{sec33}. It is important to add that in the situation the transition operator $\co$ is a generator of a Lie algebra $G$, then it is more appropriate to decompose the $(m_p,m_n)$ space into irreducible representations $\Gamma$ of $G$ and use $\Gamma$
in place of $(\tmp , \tmn)$ in Eq. (\ref{eq.sdm37}). A good example is Gamow-Teller (GT) operator and it is a generator of Wigner's spin-isospin $SU(4)$ algebra \cite{Ko-03}.  

For one-body transition operators, Eq. (\ref{eq.sdm36}) can be further simplified giving Eq. (6) in \cite{KS-00} with $\rho_{\co : biv-qN}$ and $\rho_{qN}$. Then we have,
\be
\barr{c}
\l| \lan E_f \mid \co
\mid E_i \ran \r|^2 = \dis\sum_{\alpha, \beta} \l|\epsilon_{\alpha
\beta}\r|^2\;\lan n_\beta (1-n_\alpha)  \ran^{E_i}\;\dis\f{
1}{I_{qN}(E_f|q_c)}\;\times \\
\dis\int \rho_{\co : biv-qN}\l(E_i,E_f\;;\; \ce_i, \ce_f =
\ce_i - \epsilon_\beta + \epsilon_\alpha, \overline{\sigma_i}, 
\overline{\sigma_f}, \xi |q_V\r)\;\;d\ce_i \;.
\earr \label{eq.sdm38}
\ee
Here, $\alpha$ and $\beta$ denote sp states and $\epsilon_\alpha$ and $\epsilon_\beta$ are the corresponding sp energies. The $\ce_i$ ($\ce_f$) are centroids of the configurations in the initial (final) space. Similarly,$\overline{\sigma^2_i}$ and $\overline{\sigma^2_f}$ are the average configuration variances.
Also, $\xi$ is the bivariate correlation coefficient and methods for its determination are given above. Similarly, the $q$ parameters $q_c$ and $q_V$ can be determined. It is important to stress that Eq. (\ref{eq.sdm38}) involves orbit occupancies and state densities. Starting with Eq. (\ref{eq.sdm38}) and
summing over all final energies will give transition strength sums for one-body operators. It is also possible to incorporate $J$ projection in Eq. (\ref{eq.sdm38}) (see for example \cite{app5,app6,Fl-00}). 

\subsection{Numerical tests of strength sum and strength moments} 
\label{sec54}

\begin{figure}
     \centering
         \includegraphics[width=0.875\textwidth,height=0.75\textwidth]{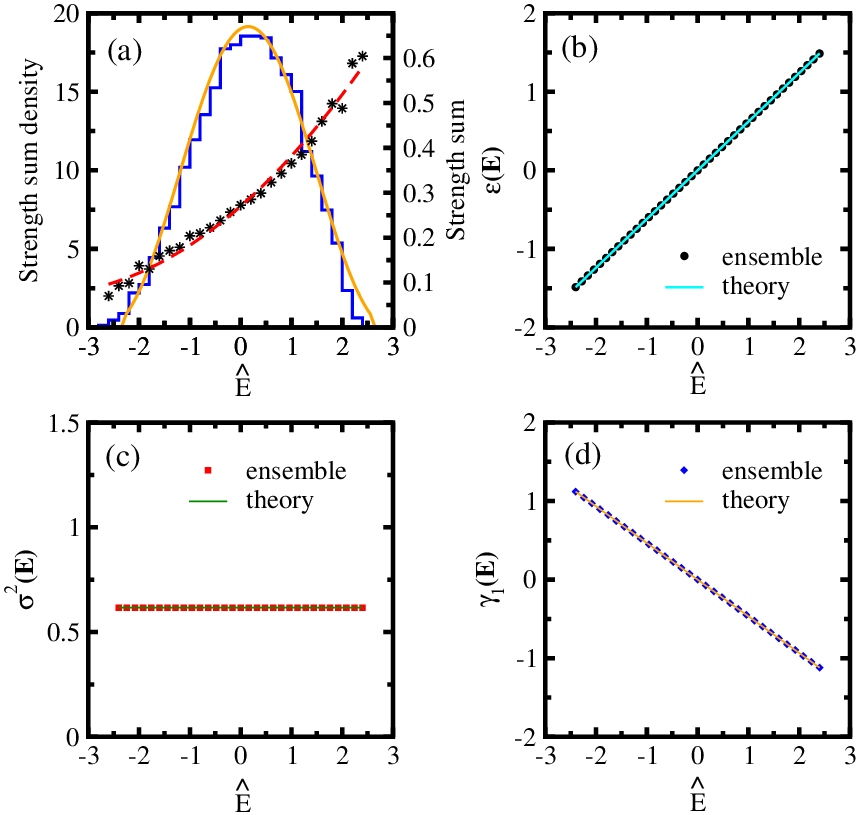}
         \caption{(a) Transition strength sum and strength density, (b) centroid $\epsilon(E)$, (c) variance $\sigma^2(E)$ and (d) skewness $\gamma_1(E)$ for a one body transition operator for an EGOE$(1+2+3)$ ensemble defined in the text. Numerical ensemble results (histogram and symbols such as stars, circles, squares and diamonds) are compared with analytical curves.}
        \label{fig:f4}
\end{figure} 

Besides transition strength sum $\lan \co^\dagger \co\ran^E$ and strength sum density $\lan \co^\dagger \co\ran^E \times I(E)$, also important are the lower order strength moments, i.e. the centroid, variance and skewness, of the distribution of strengths originating from an eigenstate with energy $E$, as they are also measurable in many situations in nuclei \cite{KK, FKPT}. For example, the moments $M_p(E)$ are given by
$$
M_p(E) = \dis\f{\dis\sum_{E_f} \l|\lan E_f \mid \co \mid E \ran\r|^2\; (E_f)^p}{\lan \co^\dagger \co \ran^E}
$$
and then the centroid is $\epsilon(E) = M_1(E)$ and the variance is $\sigma^2(E) = M_2(E) - [M_1(E)]^2$. Similarly, the skewness $\gamma_1(E)$ is defined via $M_3(E)$. As these are moments of the conditional density of the bivariate transition strength density, their variation with $E$ follows from Eqs. \eqref{eq.sdm17} and \eqref{eq.sdm18}. The $\epsilon(E)$ will be linear in $E$, the variance $\sigma^2(E)$ is a constant (does not depend on $E$) and $\gamma_1(E)$ will be linear in $E$ with negative slope. All these results are tested in a numerical example and the results are shown in Figs. \ref{fig:f4} (a)-(d). In the calculations, just as in Fig. \ref{fig:f1}, used is a EGOE$(1+2+3)$ ensemble with ($N = 12$, $m = 6$, $\lambda_2 = 0.3$ and $\lambda_3 = 0.2$). For the transition operator $\co$, chosen is the one-body operator $a^\dagger_2 a_9$. In Fig. \ref{fig:f4} (a), numerical results (histogram and stars) are compared with analytical curves for strength sum density and strength sum. They follow from Eqs. \eqref{eq.sdm32} and \eqref{eq.sdm37} with $m$-particle averages. Then, the strength sum density is a marginal of the bivariate transition strength density and thus, it follows $q$-normal distribution and this corresponds to the smooth curve in the figure. Similarly, strength sum is the ratio of strength sum density and state density. As the transition operator is not completely random and the Hamiltonian operator has a fixed one-body part along with a mixture of two and three body rank oprators, there is a shift of the centroid of the strength sum density relative to the state density centroid. Going to Figs. \ref{fig:f4} (b)-(d), it is clearly seen that the strength centroid $\epsilon(E)$, variance $\sigma^2(E)$ and skewness $\gamma_1(E)$ follow the equations for the moments of the conditional $q$-normal distribution; see  Eqs. \eqref{eq.sdm17} and \eqref{eq.sdm18} and also \cite{qMK-3}. The agreements with theory are very good with some deviations at the spectrum edges.

\section{Conclusions and future outlook}
\label{sec6}

Statistical nuclear spectroscopy (or spectral distribution method for nuclear structure studies) is till now based on the Gaussian forms for the state and transition strength densities in shell model spaces with their extension to partial densities defined over shell model subspaces. The Gaussian forms have their basis in random matrix theory with EE($k$), the embedded ensembles of $k$-body interactions in many-particle spaces. Following the recent results showing that EE($k$) in fact generate $q$-normal form for the eigenvalue densities, transition strength densities and strength functions (partial densities), in the present article developed is statistical nuclear spectroscopy based on $q$-normal (univariate and bivariate) distributions and the associated $q$-Hermite polynomials. In particular, formulation is presented for nuclear level densities in Section \ref{sec3}, for shell model orbit occupancies in Section \ref{sec4} and for transition strengths (for electromagnetic and $\beta$ and double$\beta$ type operators) and strength sums in Section \ref{sec5}. In addition, for completeness given in Section \ref{sec2} are the definition of $q$-normal distribution, $q$-Hermite polynomials and bivariate $q$-normal distribution and collected are also some of their important properties. It is important to add that the Gaussian form used in the past in statistical nuclear spectroscopy is reasonably good as long as the systems considered have sufficiently large number of particles and Hamiltonian is $(1+2)$-body. However, with growing knowledge on 3-body (perhaps also 4-body) interactions in nuclei, certainly in future one needs the formulation, with $q$-normal forms, given in Sections \ref{sec3}-\ref{sec5}. 

Going further it is necessary to carry out tests of the formulation presented in Sections \ref{sec3}-\ref{sec5} using EGOE(1+2+3) as the nuclear Hamiltonians consist of a one-body mean-field part, a stronger effective two-body part and a smaller three-body part [also a small four-body part - then we need to use EGOE(1+2+3+4)]. Results of some initial tests for state densities, orbit occupancies and transition strengths given in Sections \ref{sec3}-\ref{sec5} show that the formulation with $q$-normal forms is good. More extensive tests using EGOE(1+2+3) [also EGOE(1+2+3+4)] will be reported elsewhere along with tests using shell model codes with realistic $(1+2+3)$-body Hamiltonians. These studies, numerically intensive are expected to give insights into the role of the $q$ parameter that is related to the fourth moment, and the importance of $q$-normal distribution, that is bounded, 
in statistical nuclear physics. In addition to all these numerical tests, it is important to use the formulation given in Sections \ref{sec3}-\ref{sec5} in various applications such as  calculation of nuclear level densities, astrophysical reaction rates calculations, double $\beta$-decay transition matrix elements calculations and so on.

Before concluding, let us add that the distributions in Sections \ref{sec3}-\ref{sec5} indeed give smoothed (with respect to energy) level densities, orbit occupancies, spin-cutoff factors, transition strengths and so on. The smoothening is expected to be over a few mean spacings as fluctuations, that are neglected, operate over a few mean spacings. Level and strength fluctuations in nuclei are studied using GOE \cite{Br-81,Widen} and the role of EE in level and strength fluctuations in not yet clearly understood. This is due to the fact that even the two-point correlation function 
$$
S^{\rho}(E_i,E_f) = \overline{\rho(E_i)\,\rho(E_f)} - \overline{\rho(E_i)}\;\;\overline{\rho(E_f)}
$$
for the eigenvalues of EGOE($k$) (or EGUE($k$)) is not known \cite{Widen}. Note that the `overline' indicates ensemble average and $\overline{\rho(E)}$ is the smoothed eigenvalue density. A GC expansion with $q$-Hermite polynomials \cite{Verb4} may give a good starting point for deriving a formula for $S^{\rho}(E_i,E_f)$. It remains to be seen if the two-point function for EGOE($k$) (or EGUE($k$)) will be obtained in the near future.
 
\section{Acknowledgements}

This paper is dedicated to late Prof. J. B. French on his birth centenary. Thanks are due to N.  D.  Chavda and R.  Sahu for useful discussions and correspondence. M. V. acknowledges financial support from CONACYT project Fronteras 10872 and UNAM-PAPIIT IG101122. 


\ed